\RequirePackage{lineno}
\documentclass[aps,prd,twocolumn,superscriptaddress,showpacs,preprintnumbers,amsmath,amssymb,floatfix]{revtex4-1}
\usepackage{subfigure}
\usepackage{graphicx} 
\usepackage{dcolumn}  
\graphicspath{{ps}}

\begin{document}


\preprint{\vbox{ \hbox{   }
}}


\title{ \quad\\[1.0cm] Observation of Excited $\Omega_c$ Charmed Baryons in $e^+e^-$ Collisions}


\noaffiliation
\affiliation{University of the Basque Country UPV/EHU, 48080 Bilbao}
\affiliation{Beihang University, Beijing 100191}
\affiliation{Budker Institute of Nuclear Physics SB RAS, Novosibirsk 630090}
\affiliation{Faculty of Mathematics and Physics, Charles University, 121 16 Prague}
\affiliation{University of Cincinnati, Cincinnati, Ohio 45221}
\affiliation{Deutsches Elektronen--Synchrotron, 22607 Hamburg}
\affiliation{University of Florida, Gainesville, Florida 32611}
\affiliation{Gifu University, Gifu 501-1193}
\affiliation{SOKENDAI (The Graduate University for Advanced Studies), Hayama 240-0193}
\affiliation{Gyeongsang National University, Chinju 660-701}
\affiliation{Hanyang University, Seoul 133-791}
\affiliation{University of Hawaii, Honolulu, Hawaii 96822}
\affiliation{High Energy Accelerator Research Organization (KEK), Tsukuba 305-0801}
\affiliation{J-PARC Branch, KEK Theory Center, High Energy Accelerator Research Organization (KEK), Tsukuba 305-0801}
\affiliation{IKERBASQUE, Basque Foundation for Science, 48013 Bilbao}
\affiliation{Indian Institute of Science Education and Research Mohali, SAS Nagar, 140306}
\affiliation{Indian Institute of Technology Bhubaneswar, Satya Nagar 751007}
\affiliation{Indian Institute of Technology Guwahati, Assam 781039}
\affiliation{Indian Institute of Technology Hyderabad, Telangana 502285}
\affiliation{Indian Institute of Technology Madras, Chennai 600036}
\affiliation{Indiana University, Bloomington, Indiana 47408}
\affiliation{Institute of High Energy Physics, Chinese Academy of Sciences, Beijing 100049}
\affiliation{Institute of High Energy Physics, Vienna 1050}
\affiliation{University of Mississippi, University, Mississippi 38677}
\affiliation{INFN - Sezione di Napoli, 80126 Napoli}
\affiliation{INFN - Sezione di Torino, 10125 Torino}
\affiliation{Advanced Science Research Center, Japan Atomic Energy Agency, Naka 319-1195}
\affiliation{J. Stefan Institute, 1000 Ljubljana}
\affiliation{Kanagawa University, Yokohama 221-8686}
\affiliation{Institut f\"ur Experimentelle Kernphysik, Karlsruher Institut f\"ur Technologie, 76131 Karlsruhe}
\affiliation{Kennesaw State University, Kennesaw, Georgia 30144}
\affiliation{Department of Physics, Faculty of Science, King Abdulaziz University, Jeddah 21589}
\affiliation{Korea Institute of Science and Technology Information, Daejeon 305-806}
\affiliation{Korea University, Seoul 136-713}
\affiliation{Kyoto University, Kyoto 606-8502}
\affiliation{Kyungpook National University, Daegu 702-701}
\affiliation{\'Ecole Polytechnique F\'ed\'erale de Lausanne (EPFL), Lausanne 1015}
\affiliation{P.N. Lebedev Physical Institute of the Russian Academy of Sciences, Moscow 119991}
\affiliation{Faculty of Mathematics and Physics, University of Ljubljana, 1000 Ljubljana}
\affiliation{Ludwig Maximilians University, 80539 Munich}
\affiliation{Luther College, Decorah, Iowa 52101}
\affiliation{University of Maribor, 2000 Maribor}
\affiliation{Max-Planck-Institut f\"ur Physik, 80805 M\"unchen}
\affiliation{School of Physics, University of Melbourne, Victoria 3010}
\affiliation{Moscow Physical Engineering Institute, Moscow 115409}
\affiliation{Moscow Institute of Physics and Technology, Moscow Region 141700}
\affiliation{Graduate School of Science, Nagoya University, Nagoya 464-8602}
\affiliation{Nara Women's University, Nara 630-8506}
\affiliation{National Central University, Chung-li 32054}
\affiliation{National United University, Miao Li 36003}
\affiliation{Department of Physics, National Taiwan University, Taipei 10617}
\affiliation{H. Niewodniczanski Institute of Nuclear Physics, Krakow 31-342}
\affiliation{Nippon Dental University, Niigata 951-8580}
\affiliation{Niigata University, Niigata 950-2181}
\affiliation{Novosibirsk State University, Novosibirsk 630090}
\affiliation{Osaka City University, Osaka 558-8585}
\affiliation{Pacific Northwest National Laboratory, Richland, Washington 99352}
\affiliation{Panjab University, Chandigarh 160014}
\affiliation{University of Pittsburgh, Pittsburgh, Pennsylvania 15260}
\affiliation{Theoretical Research Division, Nishina Center, RIKEN, Saitama 351-0198}
\affiliation{University of Science and Technology of China, Hefei 230026}
\affiliation{Showa Pharmaceutical University, Tokyo 194-8543}
\affiliation{Soongsil University, Seoul 156-743}
\affiliation{Stefan Meyer Institute for Subatomic Physics, Vienna 1090}
\affiliation{Sungkyunkwan University, Suwon 440-746}
\affiliation{School of Physics, University of Sydney, New South Wales 2006}
\affiliation{Department of Physics, Faculty of Science, University of Tabuk, Tabuk 71451}
\affiliation{Tata Institute of Fundamental Research, Mumbai 400005}
\affiliation{Excellence Cluster Universe, Technische Universit\"at M\"unchen, 85748 Garching}
\affiliation{Department of Physics, Technische Universit\"at M\"unchen, 85748 Garching}
\affiliation{Toho University, Funabashi 274-8510}
\affiliation{Department of Physics, Tohoku University, Sendai 980-8578}
\affiliation{Earthquake Research Institute, University of Tokyo, Tokyo 113-0032}
\affiliation{Department of Physics, University of Tokyo, Tokyo 113-0033}
\affiliation{Tokyo Institute of Technology, Tokyo 152-8550}
\affiliation{Tokyo Metropolitan University, Tokyo 192-0397}
\affiliation{University of Torino, 10124 Torino}
\affiliation{Virginia Polytechnic Institute and State University, Blacksburg, Virginia 24061}
\affiliation{Wayne State University, Detroit, Michigan 48202}
\affiliation{Yonsei University, Seoul 120-749}

  \author{J.~Yelton}\affiliation{University of Florida, Gainesville, Florida 32611} 
  \author{I.~Adachi}\affiliation{High Energy Accelerator Research Organization (KEK), Tsukuba 305-0801}\affiliation{SOKENDAI (The Graduate University for Advanced Studies), Hayama 240-0193} 
  \author{H.~Aihara}\affiliation{Department of Physics, University of Tokyo, Tokyo 113-0033} 
  \author{S.~Al~Said}\affiliation{Department of Physics, Faculty of Science, University of Tabuk, Tabuk 71451}\affiliation{Department of Physics, Faculty of Science, King Abdulaziz University, Jeddah 21589} 
  \author{D.~M.~Asner}\affiliation{Pacific Northwest National Laboratory, Richland, Washington 99352} 
  \author{V.~Aulchenko}\affiliation{Budker Institute of Nuclear Physics SB RAS, Novosibirsk 630090}\affiliation{Novosibirsk State University, Novosibirsk 630090} 
  \author{T.~Aushev}\affiliation{Moscow Institute of Physics and Technology, Moscow Region 141700} 
  \author{R.~Ayad}\affiliation{Department of Physics, Faculty of Science, University of Tabuk, Tabuk 71451} 
  \author{T.~Aziz}\affiliation{Tata Institute of Fundamental Research, Mumbai 400005} 
  \author{V.~Babu}\affiliation{Tata Institute of Fundamental Research, Mumbai 400005} 
% ...
\collaboration{The Belle Collaboration}


\begin{abstract}
Using the entire Belle data sample of 980 ${\rm fb}^{-1}$ of $e^+e^-$ collisions,
we present the results of a study of excited $\Omega_c$ charmed baryons
in the decay mode $\Xi_c^+K^-$. We show confirmation of four of the five narrow states 
reported by the LHCb Collaboration: the $\Omega_c(3000)$ , $\Omega_c(3050)$,
$\Omega_c(3066)$, and
$\Omega_c(3090)$.

\end{abstract}

\pacs{14.20.Lq}

\maketitle


{\renewcommand{\thefootnote}{\fnsymbol{footnote}}}
\setcounter{footnote}{0}

The $\Omega_c^0$~\cite{CC} charmed baryon is a combination of $css$ quarks. 
Charmed baryons can be treated as a heavy ($c$) quark 
and a light (in this case $ss$) diquark~\cite{T1,T2,T3}. 
The ground state of $\Omega_c^0$ can be considered as a spin-1 
diquark in combination with the charm quark, as symmetry rules do not allow a spin-0 diquark. 
Thus, the ground-state $\Omega_c$, although weakly decaying, 
has a quark structure analagous to the $\Sigma_c$ and $\Xi_c^{\prime}$ rather than $\Lambda_c$ and $\Xi_c$ baryons. 
Until recently, the only excited state of the $\Omega_c^0$  observed was the $J=\frac{3}{2}^+$ state 
known as the $\Omega_c^{*0}$~\cite{Omcstar1,Omcstar2}, which decays electromagnetically into the ground state. 
All excitations have restricted
decay possibilities, because the decay $\Omega_c^{*0} \to \Omega_c^0\pi^0$ would violate isospin conservation. 
However, provided there is sufficient mass, strong decays into $\Xi_c\bar{K}, \Xi_c^{\prime }\bar{K}$, and $\Xi_c^{*}\bar{K}$
are possible.

Recently, the LHCb collaboration announced the discovery of five narrow resonances in the final state 
$\Xi_c^+K^-$~\cite{LHCb}. In addition they showed a wide enhancement at the higher mass of 3.188 ${\rm GeV}/c^2$, 
which may comprise more than one 
state. 
Here we present the results of an analysis of the same final state using data from the Belle experiment, 
and confirm many of the LHCb discoveries. 

This analysis uses a data sample of $e^+e^-$ annihilations recorded by the Belle detector~\cite{Belle} 
operating at the KEKB asymmetric-energy $e^+e^-$ collider~\cite{KEKB}. 
It corresponds to an integrated luminosity of 980 ${\rm fb}^{-1}$.
The majority of these data were taken with the accelerator energy tuned for production of the $\Upsilon(4S)$ resonance, as this is optimum
for investigation of $B$ decays. However, the excited charmed baryons in this analysis are 
produced in continuum charm production and are of higher momentum than those that are
decay products of $B$ mesons, so the data set used in this analysis also includes the Belle data taken at 
beam energies corresponding to the other $\Upsilon$ resonances and the nearby continuum ($e^+e^- \to q\bar{q}$, 
where $q \in \{u,\,d,\,s,\,c\}$).

The Belle detector is a large-solid-angle spectrometer comprising six sub-detectors: the Silicon Vertex Detector (SVD), the 50-layer Central
Drift Chamber (CDC), the Aerogel Cherenkov Counter (ACC), the Time-of-Flight scintillation counter (TOF),
the electromagnetic calorimeter, and the
$K_L$ and muon detector. A superconducting solenoid produces a 1.5 T magnetic field throughout the first five of these sub-detectors.
The detector is described in detail elsewhere~\cite{Belle}. 
Two inner detector configurations were used. The first comprised a 2.0 cm 
radius beampipe and a 3-layer silicon vertex detector, 
and the second a 1.5 cm radius beampipe and a 4-layer silicon detector and a small-cell inner drift chamber.

\begin{figure}[htb]                                                                                                                   
\includegraphics[width=3.5in]{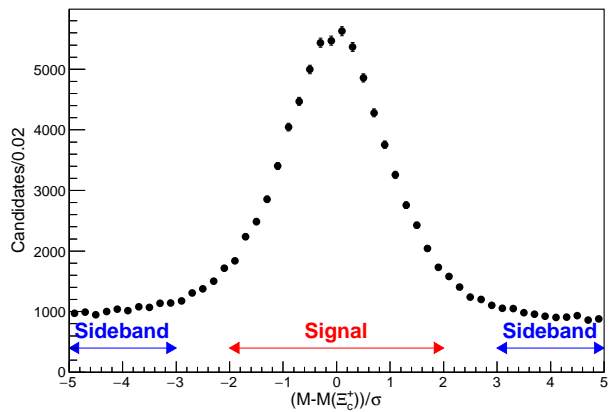}
\caption{ The distribution of the ``pull-mass'', that is $(M_{measured}-M(\Xi_c^+))/\sigma$, for all reconstructed modes
of $\Xi_c^+$ baryons. There is a requirement of $x_p > 0.65$.}
\label{fig:xic}
\end{figure}

In 2016, Belle published~\cite{Xic} the results of an analysis of excited $\Xi_c$ states 
decaying into $\Xi_c^{+/0}$ and a photon and/or pions.
To do this, seven different $\Xi_c^+$ decay modes 
($\Xi^-\pi^+\pi^+$, $\Lambda K^-\pi^+\pi^+$, $\Xi^0\pi^+$, $\Xi^0\pi^+\pi^-\pi^+$, $\Sigma^+K^-\pi^+$, $\Lambda K^0_S\pi^+$, and $\Sigma^0K^0_s\pi^+$)
were reconstructed.
The analysis presented here uses the identical reconstruction chains and the same selection 
criteria to reconstruct these same ground state $\Xi_c^{+}$ baryons.
The $\Xi_c^+$ candidates are made by kinematically fitting the decay
daughters to a common decay vertex. The position of the interaction point (IP) 
is not included in this vertex, as the small decay length associated
with the $\Xi_c^+$ decays, though very short, is not completely negligible. The
$\chi^2$ of this vertex is required to be consistent with all the daughters having a common parent.
Those combinations with a measured mass within 2 standard deviations of the nominal mass of the $\Xi_c^+$~\cite{PDG} are then constrained to that mass
and retained for further analysis. The resolution of the $\Xi_c^+$ signals depends on the decay mode and has a 
range of 3.2-15.0 ${\rm MeV}/c^2$.
In Fig.\ref{fig:xic}, we show the yield and signal-to-noise ratio of the reconstructed $\Xi_c^+$ candidates by plotting 
the ``pull-mass'', \textit{i.e.,} the difference in the reconstructed mass of the candidate and the nominal mass of the $\Xi_c^+$ 
divided by the
resolution, for all the modes together. The candidates in this distribution have a requirement on the scaled momentum, 
$x_p=p^*c/\sqrt{s/4-M^2c^4}$ of $x_p > 0.65$, where $p^*$ is the momentum of the combination in the $e^+e^-$ 
center-of-mass frame,
$s$ is the total center-of-mass
energy squared, $M$ is the invariant mass of the combination, and $c$ is the speed of light.
This requirement is not applied as part of the final analysis as we prefer to 
place an $x_p$ cut requirement only on the $\Xi_c^+K^-$ combinations;
however, it serves to display
the approximate signal-to-noise ratio of our reconstructed $\Xi_c^+$ baryons.

To investigate resonances decaying into $\Xi_c^+K^-$, $\Xi_c^+$ candidates obtained as described above are 
combined with an appropriately charged kaon candidate not
contributing to the reconstructed $\Xi_c^+$.
The kaons used to make these combinations are identified using the same criteria as in the $\Xi_c^+$ resonstruction.
That is, they are selected using the likelihood
information from the tracking (SVD, CDC) and charged-hadron identification (CDC, ACC, TOF) systems into a combined likelihood,
${\cal L}(K:h) = {\cal L}_{K}/({\cal L}_{K} + {\cal L}_{h})$ where $h$ is a proton or a pion, 
with requirements of ${\cal L}(K:p)>0.6$ and ${\cal L}(K:\pi)>0.6$. These requirements are approximately $93\%$ efficient.

To optimize the mass resolution, a vertex
constraint of the particles is made with the IP included. 
All
decay modes of the $\Xi_c^+$ are considered together. 
We then place a requirement of $x_p > 0.75$ on the $\Xi_c^+K^-$ combination.
This requirement is typical for studies of orbitally excited charmed baryons as they are known to be produced with much 
higher average momenta than the combinatorial background.

\begin{figure*}[htb]                                                                                                                   
\includegraphics[width=6.5in]{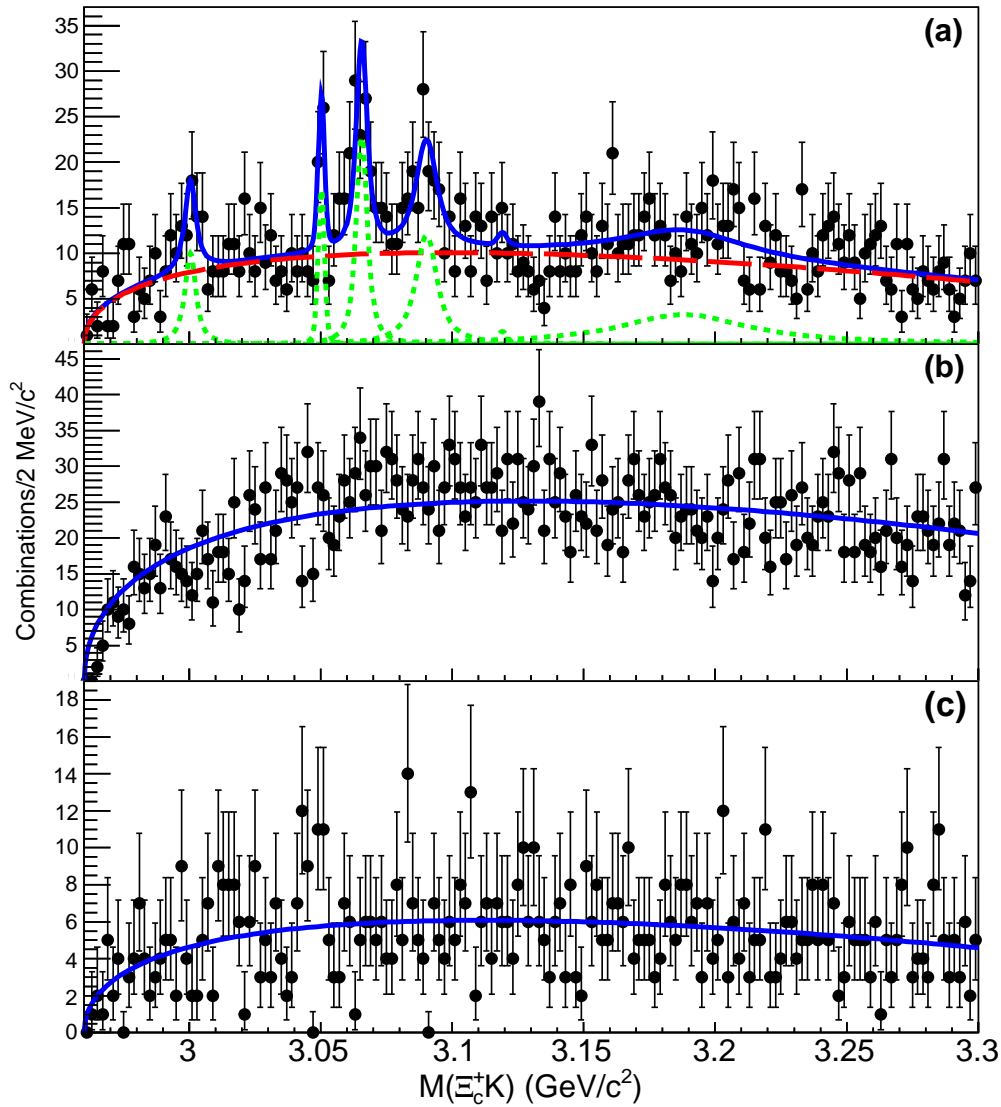}
\caption{ (a) The $\Xi_c^+ K^-$ invariant mass distribution. 
The fit shown by the solid line is the sum of a threshold function (dashed line) 
and six Voigtian (Breit-Wigner convolved with Gaussian resolution) functions, with fixed masses,
intrinsic widths and resolutions (dotted lines). (b) A threshold function fit to the $\Xi_c^{+}K^+$ (wrong-sign) 
invariant mass distribution. (c) 
A threshold function fit to the invariant mass
distribution for sidebands to the $\Xi_c^+$ candidates in combination with $K^-$ candidates. }
\label{fig:XicK}
\end{figure*}

Figure~\ref{fig:XicK}(a) shows the invariant mass distribution of the $\Xi_c^+K^-$ combinations in the mass range of interest, which 
starts at the kinematic threshold. A fit is made to this spectrum, comprising 
six signal functions and 
a background threshold
function of the form $A\sqrt{\Delta M} + B\Delta M$,
where $\Delta M $ is the mass difference from threshold, and $A$ and $B$ are free parameters. 
Each of the signal functions is a Voigtian function (a Breit-Wigner function
convolved with a Gaussian resolution). The masses and intrinsic widths of all six are fixed to the values found by 
LHCb~\cite{LHCb}. The resolutions are obtained from Monte Carlo simulation, and vary from 0.72 ${\rm MeV}/c^2$ 
for the lowest-mass peak to 1.96 ${\rm MeV}/c^2$ for the 
high-mass wide resonance. We use an unbinned likelihoood fit.
Figure~\ref{fig:XicK}(b) shows the same distribution for wrong-sign, \textit{i.e.} $\Xi_c^+K^+$ combinations.
The background function, with floating values of A and B, fits well to this distribution.
Figure~\ref{fig:XicK}(c) shows the same distribution using
$\Xi_c^+$ candidates with reconstructed masses between three and five standard deviations from the canonical mass. 
Again, this sideband distribution shows no significant
peaks, and the background function, with floating values of A and B, fits the distribution well. 

\begin{table*}[htb]
\caption{Yields of the six resonances, and comparison of the mass measurements to the LHCb values. In rows 4 and 5,
the units are MeV/$c^2$. None of the mass measurements include the uncertainty in the ground-state $\Xi_c^+$ which is common
to both experiments.}

\begin{tabular}
 {
  @{\hspace{0.1cm}}c@{\hspace{0.1cm}} | @{\hspace{0.1cm}}c@{\hspace{0.1cm}} |
  @{\hspace{0.1cm}}c@{\hspace{0.1cm}} | @{\hspace{0.1cm}}c@{\hspace{0.1cm}} |
  @{\hspace{0.1cm}}c@{\hspace{0.1cm}} | @{\hspace{0.1cm}}c@{\hspace{0.1cm}} | 
  @{\hspace{0.1cm}}c@{\hspace{0.1cm}}
}

\hline \hline
$\Omega_c$ Excited State & 3000  & 3050 & 3066 & 3090 & 3119 & 3188  \\

 Yield       & $37.7\pm 11.0$ & $ 28.2\pm 7.7 $ & $81.7 \pm 13.9$  &  $ 86.6\pm 17.4$ & $ 3.6\pm 6.9$ & $135.2\pm43.0$        \\
 Significance & 3.9$\sigma$ & 4.6$\sigma$ & 7.2$\sigma$ & 5.7$\sigma$ & 0.4$\sigma$ & 2.4$\sigma$ \\
\hline
LHCb Mass     & $3000.4\pm 0.2\pm 0.1$  & $3050.2\pm0.1\pm0.1$ & $3065.5\pm 0.1\pm0.3$ & $3090.2\pm 0.3\pm0.5$ & $3119\pm0.3\pm0.9$ &$3188\pm5\pm13$ \\
Belle Mass  & $3000.7\pm 1.0\pm0.2$  & $3050.2\pm0.4\pm0.2$ & $3064.9\pm0.6\pm0.2$ & $3089.3\pm1.2\pm0.2$ &-&$3199\pm9\pm4$ \\                 
 (with fixed $\Gamma$) &   &  &  &  & & \\                 
 
\hline
\hline

\end{tabular}

\label{tab:Table1}
\end{table*}

Table~\ref{tab:Table1} shows the yield for each of the five narrow resonances and the wide enhancement reported by LHCb. 
The significance of each signal is calculated by excluding that one peak from the fit, finding the change in the log-likelihood 
($\Delta[log(L)]$), 
and expressing the significance in terms of standard deviation using the formula $n_{\sigma} = \sqrt{2\Delta[log(L)]}$. 
Systematic uncertainties are included by calculating the signficances using a series of different fits
and choosing the lowest resultant significance value.
The differences in the fits considered are the use of different masses and widths
within the uncertainties
of the LHCb result, allowing the presence or not of an extra $C\Delta M^2$ term in the threshold function,
changing the functions fitting the peaks from Voigtian functions to s-wave relativistic Breit-Wigner functions convolved with
the resolution functions, and lastly adding or not extra functions representing possible feed-down from $\Omega_c(3066)$,
$\Omega_c(3080)$ and $\Omega_c(3119)$ decays to $\Xi_c^{\prime +}K^-$ as seen by LHCb, with shapes found 
by Monte Carlo simulation,
and floating yields.

It is clear that these data unambiguously confirm the existence of the $\Omega_c(3066)$ and $\Omega_c(3090)$. Signals of 
reasonable significance are seen for the 
$\Omega_c(3000)$ and the $\Omega_c(3050)$, but no signal is apparent for the $\Omega_c(3119)$. 
We note that, for the four narrow signals seen, we find the ratio of yields with respect to LHCb to be $\approx 0.036$. If this were
also to hold for the $\Omega_c(3119)$, we would expect 
an $\Omega_c(3119)$ signal yield of $\approx 17$, whereas we find $3.6\pm6.9$. 
Thus our non-observation of this particle is not in disagreement with LHCb.   
There is an excess in the Belle data around $3.188\ {\rm GeV}/c^2$, which 
may (as was the case in the LHCb data) be due to one or more particles.

We can measure the masses of the five confirmed signals, by fitting the same distribution without constraining the masses. 
In all cases, the masses
we find are consistent with the LHCb values, as shown in 
Table~\ref{tab:Table1}. The systematic uncertainty in the reconstruction 
of these masses is smaller than the statistical uncertainties. The uncertainty due to the knowledge of the momentum scale
is less than 0.05 ${\rm MeV}/c^2$, which is small compared with the other uncertainties. 
The systematic uncertainties in Table~\ref{tab:Table1} are dominated by the variations of the measured masses when
fitting with different values of the intrinsic widths as defined by the uncertainties in the LHCb measurements,
and the use of different --- yet reasonable --- background 
functions in the fit as was done when calculating the significances of the 
signals. 
In addition to the uncertainties shown in Table~\ref{tab:Table1}, 
there is an important systematic uncertainty of ($+0.3,-0.4$) 
${\rm MeV}/c^2$ common to the Belle and LHCb mass measurements, due to the mass measurement
of the ground state $\Xi_c^+$~\cite{PDG}.

Five states, each with one unit of orbital angular momentum between
the diquark and the charm quark, are naturally predicted by the {heavy-quark}---{light-diquark} model of baryons~\cite{T1}.
Since the LHCb observation, there have been several theoretical interpretations of the five narrow states 
found~\cite{T2a,T2b,T2c,T2d,T2e},
either in terms of these five states or by other configurations of the quarks. 
The wide state at higher mass appears to fit the pattern of wide states at around 500 ${\rm MeV}/c^2$ 
above the ground-state
charmed baryons (the $\Lambda_c^+(2765)$ and  $\Xi_c^{+/0}(2970)$). A possible explanation is that they
are the radial excitations of the ground state, with $J^P=\frac{1}{2}^+$.

To conclude, of the five narrow resonances 
observed in the $\Xi_c^+K^-$ mass spectrum by LHCb, we strongly confirm the $\Omega_c(3066)$ and $\Omega_c(3090)$ 
with very similar parameters and
confirm two more --- the $\Omega_c(3000)$ and $\Omega_c(3050)$ --- with less significance, but cannot confirm the $\Omega_c(3119)$. 
In addition, we present indications that there is wide excess, consistent with that found by LHCb, at higher mass.

We thank the KEKB group for the excellent operation of the
accelerator; the KEK cryogenics group for the efficient
operation of the solenoid; and the KEK computer group,
the National Institute of Informatics, and the 
PNNL/EMSL computing group for valuable computing
and SINET5 network support.  We acknowledge support from
the Ministry of Education, Culture, Sports, Science, and
Technology (MEXT) of Japan, the Japan Society for the 
Promotion of Science (JSPS), and the Tau-Lepton Physics 
Research Center of Nagoya University; 
the Australian Research Council;
Austrian Science Fund under Grant No.~P 26794-N20;
the National Natural Science Foundation of China under Contracts 
No.~10575109, No.~10775142, No.~10875115, No.~11175187, No.~11475187, 
No.~11521505 and No.~11575017;
the Chinese Academy of Science Center for Excellence in Particle Physics; 
the Ministry of Education, Youth and Sports of the Czech
Republic under Contract No.~LTT17020;
the Carl Zeiss Foundation, the Deutsche Forschungsgemeinschaft, the
Excellence Cluster Universe, and the VolkswagenStiftung;
the Department of Science and Technology of India; 
the Istituto Nazionale di Fisica Nucleare of Italy; 
National Research Foundation (NRF) of Korea Grants No.~2014R1A2A2A01005286, No.~2015R1A2A2A01003280,
No.~2015H1A2A1033649, No.~2016R1D1A1B01010135, No.~2016K1A3A7A09005603, No.~2016R1D1A1B02012900; Radiation Science Research Institute, Foreign Large-size Research Facility Application Supporting project and the Global Science Experimental Data Hub Center of the Korea Institute of Science and Technology Information;
the Polish Ministry of Science and Higher Education and 
the National Science Center;
the Ministry of Education and Science of the Russian Federation and
the Russian Foundation for Basic Research;
the Slovenian Research Agency;
Ikerbasque, Basque Foundation for Science and
MINECO (Juan de la Cierva), Spain;
the Swiss National Science Foundation; 
the Ministry of Education and the Ministry of Science and Technology of Taiwan;
and the U.S.\ Department of Energy and the National Science Foundation.


\begin{thebibliography}{99}
\bibitem{CC}
Throughout this paper, 
the inclusion of the charge-conjugate mode decay is implied
unless stated otherwise.
\end{thebibliography}
\end{document}